\documentclass[twocolumn,superscriptaddress,amssymb,aps,pra]{revtex4}
\usepackage{amsmath,amsfonts,amssymb,amsthm,graphics,graphicx,epsfig,bbm}
\usepackage[colorlinks=true,citecolor=blue,linkcolor=blue,urlcolor=blue]{hyperref}
\usepackage[usenames]{color}

\usepackage{dsfont}
\usepackage{txfonts}
\usepackage{amsfonts}
\usepackage{mathrsfs}
\usepackage{physics}
\usepackage{subfigure}
\usepackage{color}
\usepackage{epstopdf}
\usepackage{bbold}

\begin{document}

\title{Implementation of hybridly protected quantum gates}

\author{Chunfeng Wu}
\email{chunfeng\_wu@sutd.edu.sg}
\affiliation{Science, Mathematics and Technology, Singapore University of Technology and Design, 8 Somapah Road, Singapore 487372, Singapore}

\author{Chunfang Sun}
\affiliation{Center for Quantum Sciences and School of Physics, Northeast Normal University, Changchun 130024, China}

\author{Gangcheng Wang}
\affiliation{Center for Quantum Sciences and School of Physics, Northeast Normal University, Changchun 130024, China}

\author{Xun-Li Feng}
\affiliation{Department of Physics, Shanghai Normal University, Shanghai 200234, China}

\author{Xuexi Yi}
\affiliation{Center for Quantum Sciences and School of Physics, Northeast Normal University, Changchun 130024, China}

\begin{abstract}
We explore the implementation of hybridly protected quantum operations combining the merits of holonomy, dynamical decoupling approach and dephasing-free feature based on a simple and experimentally achievable spin model. The implementation of the quantum operations can be achieved in different physical systems with controllable parameters. The protected quantum operations are hence controllable, well-suited for resolving various quantum computation tasks, such as executing quantum error-correction codes or quantum error mitigation. Our scheme is based on experimentally achievable Hamiltonian with reduced requirement of computational resources and thus, it brings us closer towards realizing protected quantum operations for resolving quantum computation tasks in near-term quantum devices.

\end{abstract}

\date{\today}

\maketitle

In recent years, quantum computation has demonstrated its super capabilities partially with the state-of-art quantum technologies, however, we are still in the early stage of realizing full potential of quantum computation. Different types of errors in quantum system evolution are one of the main issues hindering the further development of quantum computation. In principle, the errors could be detected and corrected by adding more qubits according to quantum error-correction codes (QECCs) \cite{shor96,aharnov97,NC,knill05}. Unfortunately, it is challenging to achieve required threshold values of errors with moderate computational resources in practical experiments, for realizing large-scale fault-tolerant quantum computation \cite{Kitaev97,Ben98,Zurek98}. Thereafter introduces the theory of quantum error mitigation (QEM) for improving the accuracy of quantum algorithms with noisy intermediate-scale quantum devices \cite{Li18,Kim20}, though the performance of the QEM also relies on the reliability of quantum operations at the level of physical qubits. It is therefore of essential importance to achieve high-fidelity quantum operations for the development of quantum computation.

Various methods aiming to protect quantum operations have been investigated in the literature, such as holonomic quantum gates \cite{berry1984,aharonov87,paolo1999,sjoqvist12,mousolou2014njp}, dynamical decoupling (DD) approach \cite{dd1,dd2,dd3} and the decoherence-free subspaces (DFSs) \cite{dfs1,dfs2,dfs3}, etc. Quantum holonomy or geometric phase results in robust quantum operations insensitive to control imprecisions \cite{berry1984,aharonov87,paolo1999,sjoqvist12,mousolou2014njp}.   
DD is one attractive approach to combat errors by using external controls to approximately average out undesired couplings with relatively modest resources \cite{dd1,dd2,dd3}. DFSs provide efficient error-prevention schemes for fighting decoherence since they are the subspaces invariant to non-unitary dynamics \cite{dfs1,dfs2,dfs3}. The preserved quantum operations by holonomy, and/or DFS, and/or DD are apparently useful in mitigating errors at the level of physical qubits, and hence have gained considerable attention in applying them in different quantum protocols, for example, protected QECCs \cite{lidar09,lidar09pra,brun15,Nori18,Chen20}. By combining QECCs and the noise-resistant methods, it is expected the effect of errors can be further diminished without the need of a big increase in computational resource. The integration of QECCs and holonomic quantum operations has been explored in Refs. \cite{lidar09,lidar09pra,brun15,Nori18,Chen20}, but the integration of QECCs and multiple noise-resistant approaches is still open.

Actually there have been various schemes proposed for achieving robust quantum operations involving more than one of the noise-resistant methods, in order to mitigate the effects of multiple types of errors \cite{Kwek12,Long14,Wang15,Xue16,Zhao20}. Specifically in Refs. \cite{Kwek12,Long14,Wang15,Zhao20}, two noise-resistant methods are integrated and in Ref. \cite{Xue16}, three methods are consolidated to achieve robust quantum operations. Nevertheless, the results in the literature are not effortlessly useful for implementing various quantum computation tasks. Some of the reasons are discussed as follows. First, when combining the different noise-resistant methods, only some specific quantum gate operations can be implemented and sometimes the gate operations may not be readily the gates required in different quantum algorithms. Multi-step system evolution may be required in achieving desired quantum operations involved in certain quantum algorithm and as a result, it is unknown whether the gate fidelities are sufficiently high without evaluation. Secondly, a large number of physical qubits are required due to the encoding of physical qubits in favour of the DFS or DD approach. For examples, three physical qubits are needed to form one computational qubit in Refs. \cite{Kwek12,Long14,Wang15,Zhao20}. Thirdly, some of the required system interactions for achieving the robust quantum operations are not easily realizable in physical systems with controllable parameters given a moderate resource of external drivings \cite{Kwek12,Long14,Xue16,Zhao20}. 

In this work, we explore the implementation of hybridly protected quantum operations based on one simple spin model, combining the merits of holonomy, DD approach and dephasing-free feature. In the scheme, only two physical qubits are needed for encoding one computational qubit to support the DFS and DD approaches, reducing the requirement of computational resource. The simple spin model can be easily realized in different physical systems, and so can the hybridly protected quantum operations. We take a superconducting system as an example to explore the performance of the protected gates. With the robust quantum gates, it is possible to implement protected quantum algorithms, such as protected QECCs. Due to the noise-resistant properties offered by holonomy, DD and DFS, quantum operations are less sensitive to errors and therefore our scheme paves a promising way towards realizing fault-tolerant quantum tasks in near-term quantum devices.

To achieve hybridly protected quantum gates, we consider the following interaction,
\begin{eqnarray}\label{model}
H=\sum_{m,n\neq m}\Omega_{mn}\big(\sigma_+^m\sigma_-^n+\sigma_-^m\sigma_+^n\big),
\end{eqnarray}
where $\Omega_{mn}$ describes the coupling strength between qubits $m$ and $n$, and $\sigma_{\pm}^{m,n}=\frac{1}{2}(\sigma_x^{m,n}\pm i\sigma_y^{m,n})$. The type of interaction has been well explored in different physical systems with the state-of-art experimental technologies, such as superconducting circuits \cite{Tsai07,Nori08,Nori10,Houck11,Semba17,Oliver18,Wu20} and trapped ions \cite{Molmer99,Monroe00,Monroe17} with controllable system parameters. In most of the cases, the interaction can be readily achieved on neighbour pairs of physical qubits by applying or removing external drivings. In other words, the above desired interaction and hence the hybridly protected quantum operations discussed in the following are executable in different practical experiments.

We first use three physical qubits to implement single-qubit protected quantum gates. Qubit A is auxiliary qubit and computational qubit is formed by qubits 1 and 2 in the following way $\ket{0}_L=\ket{0}_1\ket{1}_2$ and $\ket{1}_L=\ket{1}_1\ket{0}_2$, and therefore computational qubits are encoded in the DFS of dephasing. Given a system Hamiltonian by properly applying external drivings,
\begin{eqnarray}\label{eff1}
H_1=\Omega_{A1}\sigma_+^A\sigma_-^1+\Omega_{A2}\sigma_+^A\sigma_-^2+h.c.,
\end{eqnarray}
at $\tau_1=\pi/\Omega$ with $\Omega=\sqrt{\Omega_{A1}^2+\Omega_{A2}^2}$ and $\tan\frac{\theta}{2}=\Omega_{A1}/\Omega_{A2}$, we have an evolution operator from $H_1$,
\begin{eqnarray}
U_1(\tau_1)=\ket{0}_A\bra{0}\otimes U_{1,0}(\tau_1)+\ket{1}_A\bra{1}\otimes U_{1,1}(\tau_1),
\end{eqnarray}
where $U_{1,m}(\tau_1)=(-1)^{m+1}\cos\theta\sigma^L_z-\sin\theta\sigma^L_x$ with $m=0,1$, written in the basis of $\{\ket{0}_L,\ket{1}_L\}$. If the auxiliary qubit is initially in its ground state $\ket{1}_A$, we obtain $U_{1,1}(\tau_1)$ acting on the computational qubit. We then define two subspaces $\mathcal{S}_0=\{\ket{001},\ket{010}\}$ and $\mathcal{S}_1=\{\ket{101},\ket{110}\}$ with 
the corresponding projection operators $P_{\mathcal{S}_0}=\ket{001}\bra{001}+\ket{010}\bra{010}$ and $P_{\mathcal{S}_1}=\ket{101}\bra{101}+\ket{110}\bra{110}$, to show the quantum holonomy possessed by $U_{1,1}(\tau_1)$. We first find $\mathcal{S}_m(\tau_1)=U_1(\tau_1)\mathcal{S}_m=\mathcal{S}_m$, which is first condition for quantum holonomy. Next we observe $P_{\mathcal{S}_m}(t)H_1P_{\mathcal{S}_m}(t)=0$ where $P_{\mathcal{S}_m}(t)=U_1(t)P_{\mathcal{S}_m}U^{\dagger}_1(t)$. Therefore $U_{1,1}(\tau_1)$ is a holonomic matrix in the DFS of dephasing.

We then adjust system parameters to obtain the following Hamiltonian,
\begin{eqnarray} \label{eff2}
H_2=\Omega_{12}\sigma_+^1\sigma_-^2+h.c.,
\end{eqnarray}
and we find
\begin{eqnarray}
U_{2}(\tau_2)&=&e^{i\pi/(4\Omega_{12})H_2}U_{1}(\tau_1)U_{1}(\tau_1)\big|_{\theta=0}e^{-i\pi/(4\Omega_{12})H_2}\nonumber\\
&=&\ket{0}_A\bra{0}\otimes U_{2,0}(\tau_2)+\ket{1}_A\bra{1}\otimes U_{2,1}(\tau_2),
\end{eqnarray}
where $\tau_2=2\tau_1+\pi/(2\Omega_{12})$ and $U_{2,m}(\tau_2)=e^{i(-1)^m\theta\sigma^L_z}$.
When the auxiliary qubit is in its ground state $\ket{1}_A$, we achieve single-qubit gates $U_{2,1}(\tau_2)$ acting on the computational qubit. From $U_{2,1}(\tau_2)$, we find that $\theta$ is the total phase accumulated by the computational state $\ket{m}_L\;(m=0,1)$ and the phase is purely geometric. This is because the dynamic phase is zero from the investigation of the parallel transport condition \cite{supple}.
Therefore $U_{2,1}(\tau_2)$ describes single-qubit geometric gates acting on the computational qubit in the DFS of dephasing.

For two-qubit protected quantum gates, we consider four physical qubits to form two computational qubits in the DFS of dephasing $\{\ket{00}_L,\ket{01}_L,\ket{10}_L,\ket{11}_L\}$ where $\ket{00}_L=\ket{0}_1\ket{1}_2\ket{0}_3\ket{1}_4$ for example and auxiliary qubits are not required. The system Hamiltonian is given by, 
\begin{eqnarray}
H_3=\Omega_{23}\sigma_+^2\sigma_-^3+\Omega_{24}\sigma_+^2\sigma_-^4+h.c..
\end{eqnarray}
We next control system parameters and get 
\begin{eqnarray}
H_4=\Omega_{34}\sigma_+^3\sigma_-^4+h.c.
\end{eqnarray}
Similarly through a four-step evolution according to $H_3$ and $H_4$, at $\tau'_2=2\tau'_1+\pi/(2\Omega_{34})$ where $\tau_1'=\pi/\Omega'$ with $\Omega'=\sqrt{\Omega_{23}^2+\Omega_{24}^2}$ and $\tan\frac{\theta'}{2}=\Omega_{23}/\Omega_{24}$, we find
\begin{eqnarray}
U_4(\tau_2')&=&e^{i\pi/(4\Omega_{34})H_4}U_3(\tau_1')U_3(\tau'_1)\big|_{\theta'=0}e^{-i\pi/(4\Omega_{34})H_4}\nonumber\\
&=&e^{-i\theta'\sigma_z^L\otimes\sigma_z^L},
\end{eqnarray}
where $U_3(\tau_1')$ is the evolution operator from Hamiltonian $H_3$ at $t=\tau_1'$ written in the basis of $\{\ket{00}_L,\ket{01}_L,\ket{10}_L,\ket{11}_L\}$ \cite{supple}. 
It is clear that $\theta'$ is the total phase collected by state $\ket{mn}_L\;(m,n=0,1)$. The phase is entirely geometric since the dynamic phase is zero according to the parallel transport condition \cite{supple}.
This shows $U_4(\tau_2')$ leads to two-qubit geometric gates acting on the two computational qubits.

We have shown that a set of holonomic quantum gates acting in the DFS of dephasing can be implemented based on the interaction (\ref{model}). Special spotlight should be on another robust merit possessed by the quantum gates, and that is the feature of applicable DD approach on the system evolution without causing any disturbance. According to the DD approach, undesired couplings can be approximately averaged out by the decoupling group $\mathcal{G}$ that is usually selected as: $\mathcal{G}=\{\openone^{\otimes N},\sigma_{x}^{\otimes N},\sigma_{y}^{\otimes N},\sigma_{z}^{\otimes N}\}$ for N physical qubits~\cite{Viola1,Zanardi5}. It is easy to show in our scheme that $[H_{1,2},\mathcal{G}_{j}]=0$ when $N=3$ and $[H_{3,4},\mathcal{G}_{j}]=0$ when $N=4$. The results tell us that the resulting average system-bath coupling can roughly be reduced to nil with the decoupling group, and so the DD approach to mitigate decoherence effect is applicable in our scheme of executing holonomic quantum gates in the DFS of dephasing. Therefore, we are able to implement a set of hybridly protected quantum gates, combining the merits of holonomy, DD and DFS based on the simple and experimentally achievable interaction.

\begin{figure}[b!]
\centering
\includegraphics[width=3in]{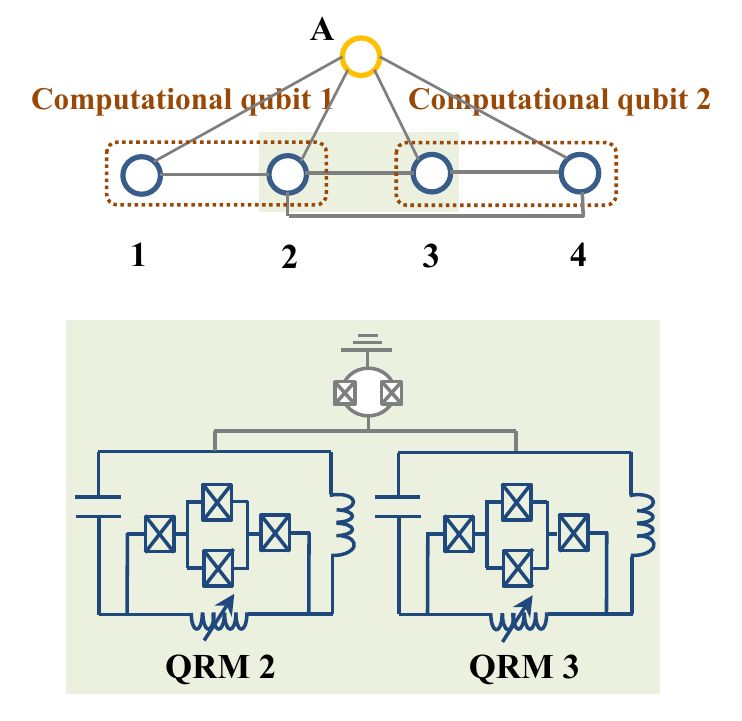}
\caption{Schematic setup of tunably coupled QRMs through SQUIDs. Each QRM includes a LC oscillator and a flux qubit, coupled to each other through inductance. QRM $A$ is auxiliary qubit to realize single-qubit gate on computational qubit 1 or 2.} \label{fig1}
\end{figure}

In the following, we investigate the implementation of the hybridly protected quantum gates in superconducting circuits. The system consists of an array of quantum Rabi model (QRM), in which each QRM is just one effective qubit and the qubit-qubit coupling is by controllable hopping interaction via superconducting quantum
interference device (SQUID), as illustrated in Fig. \ref{fig1}. The system Hamiltonian of three-neighbour QRMs is \cite{Wu20},
\begin{eqnarray}
H_{\rm cQRM} & = \sum_{m=1}^3 H^r_m + J_{12}\cos(\omega_{12}t+\phi_{12}) ( a_1^\dagger + a_1) (a_2^\dagger + a_2 )\nonumber\\
&+J_{13}\cos(\omega_{13}t+\phi_{13}) ( a_1^\dagger + a_1) (a_3^\dagger + a_3 ),
\end{eqnarray}
where $H^r_m=\omega_m^c a_m^{\dagger}a_m+\frac{\omega^q_m}{2} \sigma_{z,m}+g_m(a_m^{\dagger}+a_m)\sigma_{x,m}$ ($\omega_m^c$, $\omega^q_m$, and $g_m$ represent LC resonator frequency, flux-qubit frequency, and qubit-resonator coupling strength, respectively) is the Hamiltonian of $m$-th QRM, and $J_{mn}$, $\omega_{mn}$ and $\phi_{mn}$ describe the coupling strength, frequency and phase of the hopping interaction \cite{Wu20}. As illustrated in the supplemenary materials \cite{supple}, $H_{\rm cQRM}$ in the interaction picture with respect to $\sum_{m=1}^3 H^r_m$ almost ideally brings us the interaction described by Eq. (\ref{model}) with $\sigma_{\pm}^m$ replaced by Pauli operations acting on effective qubits \cite{Wu20}. The two desired effective Hamiltonians for realizing different protected single/two-qubit quantum gates can be derived with only one [for Hamiltonian (\ref{eff2})] or two [for Hamiltonian (\ref{eff1})] external drivings, reducing the need for more resource of external drivings. In the superconducting system, we evaluate the performance of our scheme by randomly selecting 30 initial states for each gate. The fidelities of various single- and two-qubit gates are demonstrated in Table \ref{t1}. In the calculations, five lowest energy levels of QRMs are considered and as explained in \cite{supple}, the coupled QRMs can be described by coupled effective qubits even with more higher energy levels presented by properly selecting system parameters. It is reasonable to expect poorer gate fidelities when decoherence effects are considered. Fortunately, some of the negative effects can be mitigated via the DD approach since our effective Hamiltonians commute with the decoupling group. To realize the DD approach, the decoupling group elements should be described by the Pauli operations acting on the effective qubits or QRMs. It is shown in Ref. \cite{Wu20}, single-effective-qubit operations and thus the Pauli operations can be executed by applying external drivings on physical qubits.

\begin{table}
  \begin{tabular}{ c | c | c }
    \hline\hline
Gate & Realization & Fidelity   \\ \hline
$\sigma_x^L$ & $-U_{1,1}(\tau_1)\big|_{\theta=\pi/2}$  & $\approx 0.9999+$  \\ \hline
$\sigma_z^L$ & $U_{1,1}(\tau_1)\big|_{\theta=0}$  & $\approx 0.9999+$\\ \hline
$H^L$ & $U_{1,1}(\tau_1)\big|_{\theta=-\pi/4}$  & $\approx 0.9999+$  \\ \hline
$S^L=\text{diag}(1,i)$ & $e^{i\frac{\pi}{4}}U_{2,1}(\tau_2)\big|_{\theta=\pi/4}$  &$\approx 0.9999+$  \\ \hline
$T^L=\text{diag}(1,e^{i\frac{\pi}{4}})$ & $e^{i\frac{\pi}{8}}U_{2,1}(\tau_2)\big|_{\theta=\pi/8}$  &$\approx 0.9999+$ \\ \hline
$e^{i\frac{\pi}{4}\sigma_z^L\otimes\sigma_z^L}$ & $U_{4}(\tau'_2)\big|_{\theta'=-\pi/4}$  &$\approx 0.9996$\\ \hline
$CZ^L$ & $e^{-i\frac{\pi}{4}}(S^L\otimes S^L)U_{4}(\tau'_2)\big|_{\theta'=-\pi/4}$  &$\approx 0.9994$ \\ \hline
\hline
  \end{tabular}
  \caption{The realization of a universal set of quantum gates in the system of coupled QRMs. In our calculations, the following parameters are selected, $\omega_m^c=2\pi \times 7 \;\text{GHz}$, $\omega_m^q=2\pi \times 6.1 \;\text{or}\; 5.1\; \text{GHz}$, $g_m=2\pi \times 2\;\text{GHz}$, $\Omega=\Omega'=2\pi \times 2 \;\text{MHz}$ which determine the values of $J_{mn}$, see details in supplementary materials \cite{supple}). } \label{t1}
\end{table}

We have investigated the implementation of the protected quantum gates in the superconducting circuits, both Clifford and non-Clifford gates. It is well-known that Clifford gates alone are not universal, but together with a non-Clifford $T$ gate, any quantum operations can be generated. Therefore, the set of protected quantum gates can achieve a universal set of quantum gates for quantum computation, leading to various applications in resolving quantum computation tasks. We then discuss the use of the protected quantum operations in QECCs by considering surface codes as an example. It is generally accepted that surface codes offer a promising way for achieving large-scale fault-tolerant quantum computation, partly because the syndrome measurements are based on nearest-neighbour interactions \cite{Cleland12,Meter12}. The syndrome measurements can be performed by acting Hadamard and CNOT gates in sequence on measurement and data qubits, followed by certain measurements \cite{Cleland12}. The initializtion and quantum operations on surface codes depend on established syndrome measurements aided by desired quantum operations acting on computational qubits \cite{Meter12}. Given the universal set of protected quantum gates explored in our scheme, both the initialization and quantum operations on surface codes can be preserved by holonomy, DFS and DD. Moreover, the protected quantum gates are executable in superconducting systems or trapped ions, and so it is possible to implement protected surface codes experimentally.

To summarize, we have presented a scheme to achieve protected quantum operations in a hybrid manner integrating holonomy, DD and dephasing-free features. Our scheme is based on one simple and well-developed interaction model, and in the scheme only two physical qubits are needed for encoding one computational qubit in favor of the DFS and DD approaches, and thus our scheme requires comparably less resource of qubits than that in the known schemes \cite{Kwek12,Long14,Wang15,Zhao20}. Due to the noise-resistant properties of the above mentioned approaches, the protected quantum operations are less sensitive to errors and therefore our scheme is of essential importance in achieving noise-resistant quantum computation. We have explored the implementation of the protected quantum operations in a superconducting system, in which adjustable qubit-qubit interaction between neighbour pair of QRMs is possible. Two desired effective Hamiltonians can be derived with only one or two external drivings for coupling two QRMs or two pairs of three QRMs, reducing the need for more resource and hence alleviating the scaling problem with external drivings. The set of hybridly protected quantum gates are useful in implementing various quantum computation tasks due to their universality and robustness. Specifically, we have discussed the implementation of surface codes on the basis of the protected quantum operations. Since our scheme mitigates errors in system evolution, the errors suffered at the level of creating QECCs and implementing quantum operations on QECCs can be reduced, and thus the opportunity of successfully executing QECCs can be enhanced. Moreover, it has not escaped our notice that our scheme can also be executed in a physical system of trapped ions \cite{Molmer99,Monroe00,Monroe17}. The full connectivity of trapped ions makes the protected quantum operations even useful in achieving different robust quantum computation tasks. Therefore our scheme brings us closer towards realizing protected quantum operations and so fault-tolerant quantum computation with near-term quantum devices since it is based on experimentally achievable Hamiltonian.

\vspace{5pt}
C.F.S. was supported by Fundamental Research Funds for the Central Universities (Grant No. 2412019FZ040). G.C.W. was supported by Fundamental Research Funds for the Central Universities (Grant No. 2412020FZ026) and Natural Science Foundation of Jilin Province (Grant No. JJKH20190279KJ). X.X.Y. was supported by National Natural Science Foundation of China (Grant No. 11775048).

\vspace{8pt}


\begin{thebibliography}{99}
\bibitem{shor96} D. P. DiVincenzo and P. W. Shor, Phys. Rev. Lett. \textbf{77}, 3260 (1996).

\bibitem{aharnov97} D. Aharonov and M. Ben-Or, in Proc. 29th Annual ACM Symposium on the Theory of Computation (ACM Press, New York, 1997) p. 176.

\bibitem{NC} M. A. Nielsen and I. L. Chuang, Quantum Computation and Quantum Information (Cambridge University Press, Cambridge, 2000).

\bibitem{knill05} E. Knill, Nature (London) \textbf{ 434}, 39 (2005). 

\bibitem{Kitaev97} Kitaev, A. Y. Russian, Math. Surveys 52, 1191 (1997).

\bibitem{Ben98} D. Aharonov and M. Ben-Or, inProc. 29th Ann. ACM Symp. on Theory of Computing 176ACM: New York, (1998).

\bibitem{Zurek98} E. Knill, R. Laflamme, and W. H. Zurek, Proc. Roy. Soc. A 454, 365 (1998).

\bibitem{Li18} S. Endo, S. C. Benjamin, and Y. Li, Phys. Rev. X \textbf{8}, 031027 (2018).

\bibitem{Kim20} S. Zhang, Y. Lu, K. Zhang, W. Chen, Y. Li, J.-N. Zhang, and K. Kim, Nat. Commun. \textbf{11}, 587 (2020). 

\bibitem{berry1984} M. V. Berry, Proc. R. Soc. A \textbf{ 392}, 45 (1984).

\bibitem{aharonov87} Aharonov, Y. \& Anandan, J. Phys. Rev. Lett. \textbf{ 58}, 1593 (1987).

\bibitem{paolo1999} P. Zanardi and M. Rasetti, Phys. Lett. A \textbf{ 264}, 94 (1999).

\bibitem{sjoqvist12} E. Sj\"oqvist, D. M. Tong, L. M. Andersson, B. Hessmo1, M. Johansson, and K. Singh, New J. Phys. \textbf{ 14}, 103035 (2012).

\bibitem{mousolou2014njp} V. A. Mousolou, C. M. Canali, and E. Sj\"{o}qvist, New J. Phys. \textbf{ 16}, 013029 (2014).

\bibitem{dd1} L. Viola, E. Knill, and S. Lloyd, Phys. Rev. Lett. \textbf{82}, 2417 (1999).

\bibitem{dd2} D. Li, A. E. Dementyev, Y. Q. Dong, R. G. Ramos, and S. E. Barrett, Phys. Rev. Lett. \textbf{98}, 190401 (2007).

\bibitem{dd3} J. J. L. Morton, A. M. Tyryshkin, R. M. Brown, S. Shankar, B. W. Lovett, A. Ardavan, T. Schenkel, E. E. Haller, J. W. Ager, and S. A. Lyon , Nature (London) \textbf{455}, 1085 (2008).

\bibitem{dfs1} L.-M. Duan, G.-C. Guo, Phys. Rev. Lett. \textbf{79}, 1953 (1997).
\bibitem{dfs2} P. Zanardi, M. Rasetti, Phys. Rev. Lett. \textbf{79}, 3306 (1997).
\bibitem{dfs3} D. A. Lidar, I. L. Chuang, K. B. Whaley, Phys. Rev. Lett. \textbf{81}, 2594 (1998).

\bibitem{lidar09} O. Oreshkov, T. A. Brun, and D. A. Lidar, Phys. Rev. Lett. \textbf{ 102}, 070502 (2009).

\bibitem{lidar09pra} O. Oreshkov, T. A. Brun, and D. A. Lidar, Phys. Rev. A \textbf{ 80}, 022325 (2009).

\bibitem{brun15} Y.-C. Zheng and T. A. Brun, Phys. Rev. A \textbf{ 91}, 022302 (2015).

\bibitem{Nori18} J. Zhang, S. J. Devitt, J. Q. You, and F. Nori, Phys. Rev. A \textbf{97}, 022335 (2018).

\bibitem{Chen20} C. Wu, Y. Wang, X.-L. Feng, and J.-L. Chen, Phys. Rev. Appl. \textbf{13}, 014055 (2020).

\bibitem{Kwek12} G. F. Xu, J. Zhang, D. M. Tong, E. Sj\"oqvist, and L. C. Kwek, Phys. Rev. Lett. \textbf{ 109}, 170501 (2012).

\bibitem{Long14} G. F. Xu and G. L. Long, Phys. Rev. A \textbf{ 90}, 022323 (2014).

\bibitem{Wang15} Z.-Y. Xue, J. Zhou, and Z. D. Wang, Phys. Rev. A \textbf{92}, 022320 (2015).

\bibitem{Zhao20} X. Wu and P. Z. Zhao, Phys. Rev. A \textbf{102}, 032627 (2020).

\bibitem{Xue16} C. Sun, G. Wang, C. Wu, H. D. Liu, X. L. Feng, J. L. Chen, and K. Xue, Sci. Rep. \textbf{ 6}, 20292 (2016).

\bibitem{Tsai07} A. O. Niskanen, K. Harrabi, F. Yoshihara, Y. Nakamura, S. Lloyd, and J. S. Tsai, Science \textbf{316}, 723 (2007).

\bibitem{Nori08} T. Yamamoto, M. Watanabe, J. Q. You, Yu. A. Pashkin, O. Astafiev, Y. Nakamura, F. Nori, and J. S. Tsai, Phys. Rev. B \textbf{77}, 064505 (2008).

\bibitem{Nori10} J. Q. You, X.-F. Shi, X. Hu, and F. Nori, Phys. Rev. B \textbf{81}, 014505 (2010).

\bibitem{Houck11} S. J. Srinivasan, A. J. Hoffman, J. M. Gambetta, and A. A. Houck, Phys. Rev. Lett. \textbf{106}, 083601 (2011).

\bibitem{Semba17} F. Yoshihara, T. Fuse, S. Ashhab, K. Kakuyanagi, S. Saito, and K. Semba, Nat. Phys. \textbf{13}, 44 (2017).

\bibitem{Oliver18} F. Yan, P. Krantz, Y. Sung, M. Kjaergaard, D. Campbell, J. I.J. Wang, T. P. Orlando, S. Gustavsson, W. D. Oliver, Phys. Rev. Appl. \textbf{10}, 054062 (2018).

\bibitem{Wu20} Y. Wang, Y. Su, X. Chen, and C. Wu, Phys. Rev. Appl. \textbf{14}, 044043 (2020).

\bibitem{Molmer99} A. S\o{}rensen and K. M\o{}lmer, Phys. Rev. Lett. \text{82}, 1835 (1999). 

\bibitem{Monroe00} C. A. Sackett, D. Kielpinski, B. E. King, C. Langer, V. Meyer, C. J. Myatt, M. Rowe, Q. A. Turchette, W. M. Itano, D. J. Wineland, and C. Monroe, Nature \textbf{404}, 256 (2000).

\bibitem{Monroe17} J. D. Wong-Campos, S. A. Moses, K. G. Johnson, and C. Monroe, Phys. Rev. Lett. \textbf{119}, 230501 (2017).

\bibitem{supple} See Supplemental Materials for more detailed explanations.

\bibitem{Viola1} L. Viola, S. Lloyd, and E. Knill, Phys. Rev. Lett. \textbf{83}, 4888 (1999).

\bibitem{Zanardi5} P. Zanardi, Phys. Rev. A \textbf{63}, 012301 (2000).

\bibitem{Cleland12} A. G. Fowler, M. Mariantoni, J.M. Martinis, and  A. N. Cleland, Phys. Rev. A \textbf{86}, 032324 (2012).

\bibitem{Meter12} C. Horsman, A. G. Fowler, S. Devitt, and R. V. Meter, New J. Phys. \textbf{14}, 123011 (2012).



\end{thebibliography}
\end{document}